\documentclass{article}
\usepackage{epsfig}

\date{\today}

\usepackage{graphicx}
\usepackage{epsfig}
\usepackage{amsfonts}
\usepackage{amssymb}

\newcommand{\vphi}{\varphi}

\newcommand{\ee}{\end{equation}}
\newcommand{\eea}{\end{eqnarray}}
\newcommand{\be}{\begin{equation}}
\newcommand{\bea}{\begin{eqnarray}}
\newcommand{\re}[1]{(\ref{#1})}

\newcommand{\insertplot}[5]{\begin{figure}
 \hfill\hbox to 0.05in{\vbox to #5in{\vfill
 \inputplot{#1}{#4}{#5}}\hfill}
 \hfill\vspace{-.1in}
 \caption{#2}\label{#3}
 \end{figure}}
 \newcommand{\inputplot}[3]{
 \special{ps: plotfile #1}
}
\newcounter{fig}   \newcommand{\lbfig}[1]{\refstepcounter{fig}
\label{#1} }

\begin{document}

\title{
 Axially symmetric Yang-Mills--Higgs \\
  solutions in AdS spacetime}

\date{\today}

\author{{\large Olga Kichakova}$^{1}$, {\large  Jutta Kunz$^{1}$,} 
{\large Eugen Radu$^{1}$} 
and {\large Yasha Shnir}$^{2,3}$  
\\
\\ 
 {\small  $^{1}$Institut f\"ur Physik, Universit\"at Oldenburg, Postfach 2503
D-26111 Oldenburg, Germany}\\
{\small $^{2}$School of Mathematics, University of Leeds, LS2 9JT, UK
 }\\
{\small $^{3}$Department of Theoretical Physics and Astrophysics, BSU, Minsk, Belarus
 }}
\maketitle 
 
\begin{abstract}
We consider 
axially symmetric solutions of SU(2) Yang-Mills-Higgs theory
in globally AdS spacetime and a fixed Schwarzschild-AdS black hole background.
The solutions are characterized by two integers $(m,n)$ where $m$ is related to the polar angle and 
$n$ to the azimuthal angle. 
Two types of finite energy, regular configurations are considered:
 solutions with net magnetic charge $n>1$ and
monopole-antimonopole pairs and chains with zero net magnetic charge.
The configurations are endowed with an electric charge and carry also
a nonvanishing angular momentum density.

\end{abstract}

\section{Introduction}
The study of solutions to the $SU(2)$ Yang-Mills-Higgs (YMH) equations
with an adjoint representation Higgs field, 
is a subject of long standing interest. 
This model has a non-trivial vacuum structure which gives rise to a variety
of regular, non-perturbative finite energy solutions,
such as magnetic monopoles \cite{'tHooft:1974qc}, \cite{Polyakov:1974ek},
multimonopoles \cite{Rebbi:1980yi}, \cite{Ward:1981jb}
and composite configurations,
containing both monopoles and antimonopoles 
\cite{Kleihaus:1999sx}, 
\cite{Kleihaus:2003nj}, 
\cite{Kleihaus:2003xz}\cite{Kleihaus:2004is},\cite{Kleihaus:2004fh}.
Moreover, solutions which do not possess any continuous symmetry
are also known to exist \cite{Corrigan:1981fs}.
(A review of these aspects can be found in \cite{mono-book}.)
All these configurations can be generalized to include an 
electric charge  which further enriches the pattern of the solutions,
leading, in the axially symmetric case,
 to a simple relation between the angular momentum, electric charge and total magnetic charge 
\cite{VanderBij:2001nm}, 
\cite{vanderBij:2002sq}, 
\cite{Volkov:2003ew}.

When including the effects of gravity,
a branch of gravitating solutions emerges smoothly
from the corresponding flat space configurations 
(see the review \cite{Volkov:1998cc}).
Moreover, these solutions allow for black hole generalizations, a complicated
picture emerging, whose 
 details depend mainly on the presence or not of a global
magnetic charge 
\cite{Hartmann:2001ic}, 
\cite{Kleihaus:2000kv}, 
\cite{Kleihaus:2007vf},
\cite{Kleihaus:2004gm}.

The magnetic monopole and dyon solutions of the YMH system have 
enjoyed recently 
some renewed interest 
in the context of Anti-de Sitter (AdS)/conformal field theory 
(CFT) conjecture \cite{Maldacena:1997re}.
It has been argued that, if the bulk
AdS spacetime contains non-abelian magnetic monopoles,
new interesting phenomena
may result, including spontaneous breaking of translational symmetry
in the dual theory
 \cite{Bolognesi:2010nb} 
  (see also \cite{Sutcliffe:2011sr}).
The effects of the inclusion of an electric charge on the spherically symmetric magnetic monopoles
have been
 discussed in \cite{Lugo:1999fm}.
 For black hole solutions with a planar horizon topology and possessing YMH hair,
it has been shown  in \cite{Lugo:2010qq}, \cite{Lugo:2010ch}
that the dual system defined in 2+1 dimensions 
undergoes a second order phase transition and exhibits the
condensation of a composite charge operator.
Moreover, it has been suggested in  Ref. \cite{Allahbakhshi:2011nh} that the
dual field theory is generally a field theory 
with a vortex condensate.

These results have motivated us to consider the question on how
the inclusion of a negative cosmological constant
would affect the properties of the axially symmetric YMH configurations
studied so far in a Minkowski spacetime background only.
This work is intended as a preliminary study in this direction, since,
for simplicity, we are  working
in the probe limit, $i.e.$ for a fixed (Schwarzschild-)AdS geometry, without
including the effects of the backreaction on the spacetime metric.
Moreover, we are restricting ourselves 
to a  foliation of the background geometry leading to a $(2+1)$-dimensional
Einstein universe boundary metric.

\section{The model}

\subsection{The action and field equations}
We consider the action of the $SU(2)$ YMH
theory  
\be 
\label{model}
S=-\frac{1}{4\pi}\int   d^4 x\sqrt{-g}\left (
{\rm Tr} 
\bigg \{
\frac{1}{2} 
\, F_{\mu\nu}F^{\mu\nu} 
+\frac{1}{4} 
   D_\mu \Phi\, D^\mu \Phi  
+\frac{\lambda}{8}
 \left(\Phi^2 - \eta^2 \right)^2 
  \bigg \}
  \right),
\ee
 with the 
$SU(2)$ field strength tensor
$
F_{\mu\nu} = \partial_\mu A_\nu - \partial_\nu A_\mu + i e [A_\mu, A_\nu] \ ,
$
($A_\mu $ being the gauge potential),
and the covariant derivative of the Higgs field $\Phi $
in the adjoint representation
$
D_\mu \Phi = \nabla_\mu \Phi +i e [A_\mu, \Phi] \ .
$
Here $e$ denotes the  gauge coupling constant,
$\eta$ denotes the vacuum expectation value of the Higgs field,
and $\lambda$ represents the strength of the scalar coupling.

Under $SU(2)$ gauge transformations $U$,
the gauge potentials and the Higgs field transform as
\begin{equation}
\label{gtgen} 
A_{\mu}' = U A_{\mu} U^\dagger + \frac{i}{e} (\partial_\mu U) U^\dagger,~
~~\Phi' = U \Phi U^\dagger.
\end{equation} 
 
Variation of   (\ref{model}) with respect to the gauge field $A_\mu$
and the Higgs field $\Phi$
leads to the field equations of the model
\begin{eqnarray}
\label{feqA} 
  D_\mu F^{\mu\nu} 
   =\frac{1}{4} i e [\Phi, D^\nu \Phi ] \ ,
~~
 D_\mu  D^\mu \Phi 
=\lambda (\Phi^2 -\eta^2) \Phi  \ ,
\end{eqnarray}
 while the variation with respect
to the metric $g_{\mu \nu}$ yields the energy-momentum tensor 
\begin{eqnarray}
\label{Tik}
&T_{\mu\nu} =
   {\rm Tr}
   \bigg \{
      2\, \,
    ( F_{\mu\alpha} F_{\nu\beta} g^{\alpha\beta}
   -\frac{1}{4} g_{\mu\nu} F_{\alpha\beta} F^{\alpha\beta})
   + 
 \frac{1}{2}D_\mu \Phi D_\nu \Phi
    -\frac{1}{4} g_{\mu\nu} D_\alpha \Phi D^\alpha \Phi 
   -\frac{\lambda}{8}g_{\mu\nu}  (\Phi^2 - \eta^2)^2
   \bigg \}
\ .
\end{eqnarray} 
As usual, 
the nonzero vacuum expectation value of the Higgs field
breaks the non-Abelian $SU(2)$ gauge symmetry to the Abelian U(1) symmetry.
The particle spectrum of the theory then consists of a massless photon,
two massive vector bosons of mass $M_v = e\eta$,
and a massive scalar field $M_s = {\sqrt {2 \lambda}}\, \eta$.

\subsection{ The Ansatz}

\subsubsection{The metric background}
 
 All YMH solutions in this work are studied in the probe limit, in which 
 the backreaction of the matter field on the spacetime geometry is ignored.
 This approximation which is valid as long as the dimensionless coupling constant 
 $\alpha^2=4 \pi G/\eta^2$ is very small (with $G$ the Newton's constant), greatly simplifies the problem 
 but retains most of the interesting physics.
 For example,
 the nonlinear interaction between 
 gauge fields and scalars is retained; also
  the background geometry may possess a horizon.

 For the background metric, we take first the AdS$_4$ spacetime,
 written in global coordinates
 \begin{eqnarray}
\label{AdS}
 ds^2=\frac{dr^2}{1+\frac{r^2}{\ell^2}} +r^2(d\theta^2+\sin^2 \theta d\varphi^2) - (1+\frac{r^2}{\ell^2})dt^2,
\end{eqnarray}
where $r,t$ are the radial and time coordinates, respectively  (with $0\leq r<\infty$), while
$\theta$ and $\varphi$ are angular coordinates  with the usual range, parametrizing the two dimensional sphere $S^2$.
Also, $\ell$ is the AdS length scale which is fixed by the cosmological constant, $\Lambda=-3/\ell^2$.

Since we want to study the effects of an event horizon
on the YMH solutions, we shall consider as well a Schwarzschild-AdS (SAdS) black hole background,
with a line element
\begin{eqnarray}
\label{SAdS}
 ds^2=\frac{dr^2}{N(r)} +r^2(d\theta^2+\sin^2 \theta d\varphi^2) -N(r)dt^2,~~ 
 {\rm with}~~ N(r)= 1-\frac{2M}{r}+\frac{r^2}{\ell^2},
\end{eqnarray}
  $M>0$ being a parameter which fixes the ADM mass of the solution.
For $\Lambda\to 0$, one recovers the usual Schwarzschild solution.
This black hole has an event horizon at $r=r_h$, with $r_h$ the solution of the equation $N(r_h)=0$.
 However, in practice it is more convenient to take $r_h,\ell$ as input parameters, 
  the function $N(r)$ being written as
\begin{eqnarray}
\label{metricSADS3}
N(r)= \left(1-\frac{r_h}{r} \right)\left(1+\frac{1}{\ell^2}(r^2+r r_h+ r_h^2) \right).
\end{eqnarray}
With this parametrization, the mass $M$, Hawking temperature $T_H$ and event horizon
area of the SAdS black hole are given by 
 $M= {r_h(r_h^2+\ell^2)}/({2\ell^2})$, $T_H= (1/r_h+ 3r_h/\ell^2)/(4\pi)$ 
and $A_H=4\pi r_h^2$, respectively.

\subsubsection{The matter fields}
The axially symmetric YMH configurations we study in this work are less symmetric than the
backgrounds (\ref{AdS}), (\ref{SAdS}),
being invariant under the action 
of the Killing vectors $\partial/\partial \varphi$ and $\partial/\partial t$ only.
The construction of a  YMH ansatz with these symmetries has been discussed by many authors 
starting with the pioneering papers by Manton \cite{Manton:1977ht} 
and Rebbi and Rossi \cite{Rebbi:1980yi}. 

In what follows we shall use a parametrization of the general ansatz 
which was  employed in the previous studies on 
axially symmetric YMH solutions in an asymptotically flat spacetime.
This ansatz is characterized by two integers, $ m$ and  $n$, where $m$
is related to the polar angle and $n$ to the azimuthal angle \cite{Kleihaus:2003nj}
and it reads
\begin{eqnarray}
\nonumber
&&A_\mu dx^\mu
=\left( B_1 \frac{\tau_r^{(n,m)}}{2e}+ B_2 \frac{\tau_\theta^{(n,m)}}{2e} \right)dt+
\left( \frac{K_1}{r} dr + (1-K_2)d\theta\right)\frac{\tau_\vphi^{(n)}}{2e}
- n \sin\theta \left( K_3\frac{\tau_r^{(n,m)}}{2e}
                     + K_4\frac{\tau_\theta^{(n,m)}}{2e}\right) d\vphi
\ , 
\\
 \label{ansatzA}
&&
\Phi
=
 \eta \left( \Phi_1\tau_r^{(n,m)}+ \Phi_2\tau_\theta^{(n,m)}  \right) \  .
\label{ansatzPhi}
\end{eqnarray}
where the only $\varphi$-dependent terms are the $SU(2)$ matrices
$\tau_r^{(n,m)} $, $\tau_\theta^{(n,m)} $, and $\tau_\vphi^{(n)}$.
These matrices 
are defined as products of the spatial unit vectors
\begin{eqnarray}
{\hat e}_r^{(n,m)} & = & \left(
\sin(m\theta) \cos(n\vphi), \sin(m\theta)\sin(n\vphi), \cos(m\theta)
\right)\ , \nonumber \\
{\hat e}_\theta^{(n,m)} & = & \left(
\cos(m\theta) \cos(n\vphi), \cos(m\theta)\sin(n\vphi), -\sin(m\theta)
\right)\ , \nonumber \\
{\hat e}_\vphi^{(n)} & = & \left( -\sin(n\vphi), \cos(n\vphi), 0 \right)\ ,
\label{unit_e}
\end{eqnarray}
with the Pauli matrices $\tau^a = (\tau_x, \tau_y, \tau_z)$, $i.e.$
$
\tau_r^{(n,m)} =
\sin(m\theta) (\cos(n\vphi) \tau_x + \sin(n\vphi)\tau_y) + \cos(m\theta) \tau_z \ ,
$
$
\tau_\theta^{(n,m)} =
\cos(m\theta) (\cos(n\vphi) \tau_x + \sin(n\vphi)\tau_y) - \sin(m\theta) \tau_z \ ,
$
and
$
\tau_\vphi^{(n)} =
 -\sin(n\vphi) \tau_x + \cos(n\vphi)\tau_y \ .
 $
As we shall see, this choice of the SU(2) matrices has the advantage to greatly simply the 
possible boundary conditions.
The four magnetic gauge field functions $K_i$, two electric gauge functions $B_i$ and two Higgs field functions
$\Phi_i$ depend on the coordinates $r$ and $\theta$, only.
All profile functions are even or odd w.r.t.~reflection symmetry, $\theta \rightarrow \pi-\theta$.

The symmetry of the gauge field under the spacetime Killing vectors $\partial/\partial \varphi$ and $\partial/\partial t$
means that
their action  can be compensated by a suitable gauge transformation
\cite{Heusler:1996ft,Forgacs:1980zs}.  
For the time translational symmetry, we have choosen a natural gauge such that $\partial A/\partial t$=0.
However, a rotation around the $z-$axis 
can be compensated by a gauge rotation
${\mathcal{L}}_\varphi A=D\Psi,$
(with $\Psi=n\tau_z/2e$) and therefore
$
F_{\mu \varphi} =  D_{\mu}W,
$
$
D_{\varphi}\Phi= ie[W,\Phi] \nonumber,
$
where $W=A_{\varphi}-\Psi$.

 
The gauge transformation
$
U = \exp \{i \Gamma (r,\theta) \tau_\vphi^{(n)}/2\}
$
leaves the Ansatz form-invariant \cite{Brihaye:1994ib}.
Thus, to construct regular solutions we have to fix the gauge.
In this work we have used mainly the modified form of the gauge condition \cite{Kleihaus:2003nj} 
which can be applied in
the case of a (Schwarzschild-)AdS background:
 $2r N \partial_r K_1 + r K_1 \partial_r N- 2\partial_\theta K_2 = 0.$
 
With this Ansatz and gauge choice, the equations of motion reduce to a set of
eight coupled partial differential equations,
to be solved numerically subject to the set of boundary conditions
discussed below.

\subsection{Global charges and the vacuum structure}
 \subsubsection{The mass-energy, angular momentum and the magnetic and electric charges }

Let us define the total mass-energy $E$ of the solutions as the integral over 
the three dimensional space of the energy density $\rho=-T_t^t$, 
\begin{eqnarray}
E &=& -\frac{1}{4\pi}
\int T_{t}^{t}\sqrt{-g} d^{3}x
=\int_{r_0}^{\infty}dr \int_{0}^{\pi}d\theta \int_{0}^{2\pi}d\varphi \sqrt{-g}~
   {\rm Tr} \Big\{
F_{r\theta}F^{r\theta}+F_{r\varphi}F^{r\varphi}+F_{\theta \varphi}F^{\theta \varphi}
\\
\nonumber
&&-F_{rt}F^{rt}-F_{\theta t}F^{\theta t}-F_{\varphi t}F^{\varphi t}
+\frac{1}{4}(D_r\Phi D^r\Phi+D_{\theta}\Phi D^{\theta}\Phi+D_{\varphi}\Phi D^{\varphi}\Phi
-D_t\Phi D^t\Phi)+\frac{\lambda}{8} (\Phi^2 - \eta^2)^2
\Big\}
.
\end{eqnarray}
In the above relation,
one takes $r_0=0$ for solutions in AdS spacetime and $r_0=r_h$
for a SAdS black background, in which case $E$ corresponds 
to the total mass-energy of the fields
outside the horizon.

Our solutions possess also a nonvanishing 
angular momentum density, since $T^{t}_{\varphi}\neq 0$. 
The total angular momentum of the configuration is given by the integral 
\begin{equation}
\label{J}
J = \frac{1}{4\pi}
\int d^3x \sqrt{-g} T^{t}_{\varphi} 
=
\frac{1}{4\pi} \int d^3x \sqrt{-g}~  2{\rm Tr} \Big\{ 
F_{r\varphi}F^{r t} + F_{\theta\varphi}F^{\theta t} + \frac{1}{4}D_\varphi \Phi D^t \Phi \Big\}.
\end{equation}
As proven in \cite{VanderBij:2001nm},  
the angular momentum density can be written as a total derivative
in terms of Yang-Mills potentials only, 
 \begin{eqnarray}
\label{T34}
T_{\varphi}^{t}=2 {\rm Tr} \Big\{\frac{1}{\sqrt{-g}}\partial_{\mu}(WF^{\mu t}\sqrt{-g})\Big\}.
\end{eqnarray}
As a result,  the total angular momentum stored in the YMH fields  outside the horizon
can be expressed as a difference between a term at infinity and an inner boundary contribution:
\begin{eqnarray}
\label{totalJ}
J ={\cal J}({r=\infty})-{\cal J}({r=r_0}), 
\end{eqnarray}
with ${\cal J}(R)$ the field angular momentum inside a sphere of radius $R$ defined as
\begin{eqnarray}
{\cal J}(R)=
\frac{1}{4\pi}
\oint_{r=R} 2   {\rm Tr} \{WF^{\mu t} \} dS_{\mu}
=\frac{1}{2}
 \int_0^{\pi} d \theta \sin \theta^{~}
 r^2[W^{(r)}F^{rt(r)}+W^{(\theta)}F^{rt(\theta)}+W^{(\varphi)}F^{rt(\varphi)}]\bigg |_{r=R}~.
 \end{eqnarray} 

The YMH solutions may possess also magnetic and electric charges.
A gauge-invariant definition of
these quantities is found by employing the 
  Abelian 't Hooft tensor \cite{'tHooft:1974qc}
\begin{equation}
{\cal F}_{\mu\nu} = {\rm Tr} \left\{ \hat \Phi F_{\mu\nu}
- \frac{i}{2e} \hat \Phi D_\mu \hat \Phi D_\nu \hat \Phi \right\}
= \hat \Phi^a F_{\mu\nu}^a + \frac{1}{e} \epsilon_{abc}
\hat \Phi^a D_\mu \hat \Phi^b D_\nu \hat \Phi^c
\ , \label{Hooft_tensor} \end{equation}
where the Higgs field is normalized as
$|\hat \Phi|^2 = (1/2) {\rm Tr\,} \hat \Phi^2 =\sum_a (\hat \Phi^a)^2 = 1$.
Then the 't Hooft tensor yields the electric current $j_{\rm el}^\nu$, with
%
$
 \nabla_\mu {\cal F}^{\mu\nu} = -4 \pi j_{\rm el}^\nu
$
and the magnetic current  $j_{\rm m}^\nu$,
with
%
$
 \nabla_\mu {^*}{\cal F}^{\mu\nu} = 4 \pi j_{\rm m}^\nu,
$
%
where ${^*}{\cal F}$ represents the dual field strength tensor.
Then $j_{\rm el}^t$ and $j_{\rm m}^t$  correspond  to the
electric and magnetic charge densities,   respectively.
%
%
 The electric and magnetic charges $Q_e$, $Q_m$ are given by the integrals
 \begin{equation}
 Q_m=\frac{1}{4\pi}\oint_{S^2_{\infty}} {^*}{\cal F}_{\theta \varphi} d\theta d\varphi, ~~
 Q_e=\frac{1}{4\pi}\oint_{S^2_{\infty}} {\cal F}_{\theta \varphi} d\theta d\varphi.
\end{equation}

\subsubsection{The   ground states of the model}
In the presence of the Higgs field, 
the behaviour of the matter field as $r  \to \infty$,
as imposed by finite energy requirements,
is  similar to the asymptotically  flat case\footnote{Note that without a Higgs field,
the Yang-Mills-AdS solutions exhibit a very different behaviour then in the $\Lambda\to 0$
limit, see $e.g.$ \cite{Winstanley:1998sn}.
In particular, one finds finite mass, stable configurations \cite{Breitenlohner:2003qj}. }.
 The assumption that the Higgs field approaches asympototically a constant value $|\Phi|\to \eta$,
 together with the finite energy condition $D_{\mu}\Phi D^{\mu}\Phi\to 0$,
 leads to two different sets of fundamental solutions which define the ground states of the model
 (in this discussion we set $A_t=0$ and view the electric fields as excitations).

For an even value $m=2k$ of the winding number with respect to $\theta$,
the ground state of the model  corresponds to
a gauge transformed trivial solution and the magnetic charge vanishes
\begin{eqnarray}
\label{evenk}
\Phi = \eta U \tau_z U^\dagger \   , \ \ \
A_\mu =  \ \frac{i}{e} (\partial_\mu U) U^\dagger ~.
\end{eqnarray}

The situation is different for odd values $m=2k+1=1,3,\dots$ of the winding number with respect to $\theta$.
The solutions in the sector with topological charge $n$ will approach asymptotically
a ground state with
\begin{eqnarray}
\label{oddk}
\Phi  =  U \Phi_\infty^{(1,n)} U^\dagger \   , \ \ \
A_\mu = U A_{\mu \infty}^{(1,n)} U^\dagger
+\frac{i}{e} (\partial_\mu U) U^\dagger \  ,
\end{eqnarray}
where
\begin{eqnarray}
 \Phi_\infty^{(1,n)} =\eta \tau_r^{(1,n)}\ , \ \ \
A_{\mu \infty}^{(1,n)}dx^\mu =
\frac{\tau_\vphi^{(n)}}{2e} d\theta
- n\sin\theta \frac{\tau_\theta^{(1,n)}}{2e} d\vphi
\end{eqnarray}
is the  solution describing a charge $n$ multimonopole.
Note that $U = \exp\{-i k \theta\tau_\vphi^{(n)}\}$, both
for even and odd $m$.

From these asymptotic behaviours, one gets the
 magnetic charge $Q_m$ of the solutions,
\begin{equation}
Q_m = \frac{n}{2}\left[1 - (-1)^m\right]
\ , \label{P} 
\end{equation}
$i.e.$ solutions in the topologically trivial sector $m=2k$ have no magnetic charge, $Q_m=0$,
whereas solutions in the non-trivial sectors $m=2k+1$ have magnetic charge $Q_m=n$.

\subsection{The boundary conditions}
To obtain finite energy solutions
with the proper symmetries,
we must impose appropriate boundary conditions.
Without a horizon,
these boundary conditions are similar to
those used in the previous work in the flat space limit  
\cite{Kleihaus:2003nj}, \cite{Kleihaus:2005fs}.

The large $r$ behaviour of the YMH solutions is 
fixed by the requirement that the ground states (\ref{evenk}), (\ref{oddk})
are approached asymptotically.
In terms of the functions $K_1 - K_4$ and  $\Phi_1$, $\Phi_2$ these boundary
conditions read
\begin{eqnarray}
K_1 = 0,~~
K_2= 1 - m,~~K_4= \frac{\sin(m\theta)}{\sin\theta} \ ,
\label{K12infty}
\end{eqnarray}
while for $K_3$ we impose
\begin{eqnarray}
K_3= \frac{\cos\theta - \cos(m\theta)}{\sin\theta} ~~
{\rm for}~m ~{\rm odd},~~{\rm and}~~
K_3 =\frac{1 - \cos(m\theta)}{\sin\theta} ~~
{\rm for}~ m ~ {\rm even}.
\label{K3infty}
\end{eqnarray}
We further impose the following boundary conditions at infinity for the 
electric gauge field potentials and  the
Higgs field functions, respectively
\begin{eqnarray}
\label{B12infty}
B_1 =V_0,~~   B_2=0,~~\Phi_1= 1  ,~~\Phi_2 = 0,
\end{eqnarray}
where the constant $V_0$ corresponds to the electrostatic potential of the configurations.


The boundary conditions of the matter fields on the event horizon   
results from the requirement of regularity at $r=r_h$ together with an expansion
of the field variables as a power series in $(r-r_h)$.
Here one should remark that
  the numerical study of solutions in a fixed SAdS background
is simplified by introducing a new radial coordinate $\bar r=\sqrt{r^2-r_h^2}$,
such that the event horizon resides at $\bar r=0$.
(The corresponding expression of the SAdS line element 
follows directly from (\ref{SAdS}), (\ref{metricSADS3}).)
Also, the boundary conditions at the horizon take a simpler form in this case:
the magnetic functions $K_i$ and the Higgs functions  $\Phi_1,\Phi_2$ are requested to
 satisfy Neumann boundary conditions
\bea 
\label{boundary-horizon-1}
\partial_r K_1 \big|_{\bar r=0} = \partial_r K_2 \big|_{\bar r=0}  =
\partial_r K_3 \big|_{\bar r=0}=\partial_r K_4 \big|_{\bar r=0}=
\partial_r \Phi_1 \big|_{\bar r=0} = \partial_r \Phi_2 \big|_{\bar r=0}  =0~,
\eea
while the electric field functions satisfy Dirichlet boundary conditions,
\begin{equation} 
\label{boundary-horizon-2}
B_1(\bar r=0,\theta) = B_2(\bar r=0,\theta) = 0.
\end{equation}
 
The corresponding set of boundary conditions in the AdS background limit $r_h\to 0$ are more complicated
and do not result directly from (\ref{boundary-horizon-1}), (\ref{boundary-horizon-2}).
Regularity of the solutions at the origin ($r=0$)
requires  that the magnetic gauge field functions $K_i$ should satisfy
\begin{equation} 
\label{boundary-old-1}
K_1(0,\theta)= K_3(0,\theta)= K_4(0,\theta)= 0\ , \ \ \ \
K_2(0,\theta)= 1.
\end{equation}
For the electric components of the gauge field we impose instead
\begin{equation} \label{boundary-old-2}
\sin(m\theta) B_1(0,\theta) + \cos(m\theta) B_2(0,\theta) = 0,~~
 \partial_r\left[\cos(m\theta) B_1(r,\theta)
              - \sin(m\theta) B_2(r,\theta)\right] \big|_{r=0} = 0 \ ,
\end{equation}
 the Higgs field functions $\Phi_i$ satisfing a similar set of boundary conditions
\begin{equation}
\sin(m\theta) \Phi_1(0,\theta) + \cos(m\theta) \Phi_2(0,\theta) = 0 \ ,
~~
 \partial_r\left[\cos(m\theta) \Phi_1(r,\theta)
              - \sin(m\theta) \Phi_2(r,\theta)\right] \big|_{r=0} = 0 \ .
\end{equation} 

The boundary conditions along the $z$-axis
($\theta=0$ and $\theta=\pi $) are determined by the
symmetries and regularity requirements, and read
\begin{equation}
K_1 = K_3 = \Phi_2 =0,~~
\partial_\theta K_2 = \partial_\theta K_4 = \partial_\theta \Phi_1 =0,~~
\partial_\theta B_1(r,0) =0, ~~B_2(r,0) =0.
\end{equation}
Additionally, regularity on the symmetry axis requires $K_2(r,\theta=0)=K_4(r,\theta=0)$. 
We used this condition as an additional
test of the numerical output.

\section{The results}
 
Axially symmetric solutions of the YMH equations
have been extensively studied in a Minkowski spacetime background, starting with the
pioneering work  \cite{Rebbi:1980yi}.
However, relatively little is known about the AdS counterparts of the $\Lambda=0$ configurations.
 Unlike the flat spacetime case, 
in an AdS spacetime there are no analytic solutions that might be used as a guiding line.
Moreover, when a cosmological constant is included (no matter how small $|\Lambda|$ is)
no solution close to the BPS configuration can be found \cite{Lugo:1999fm}. 
Some numerical results supporting the existence of multimonopoles and monopoles-antimonopoles with $Q_e=0$ 
in an AdS background are reported in \cite{Radu:2004ys}, while a discussion ofz the basic
properties of electrically charged $m=1$ solutions was given in Ref. \cite{VanderBij:2001nm}.
However, the effects induced by the presence of the event  
 horizon have not yet been studied in the literature, 
  except for $m=1,n=1$ spherically symmetric configurations \cite{Lugo:2010ch}.
 
 Before discussing the properties of the solutions, let us mention that 
the dependence on the massive boson vector mass $M_v=e\eta$
 is removed by using the rescaling  $r\to r/(e\eta)$, $\Lambda\to \Lambda e^2\eta^2$, $r_h\to r_h/(e\eta)$ and $B_i \to B_i/(e\eta)$. Also, to simplify the picture,  we restrict our study in this work to the Prasad-Sommerfield limit $\lambda=0$.
 Thus, apart from the value of the event horizon radius $r_h$, 
 the configurations will depend on the
winding numbers $m$ and $n$, on the parameter
$V_0$ which fixes the value of the electric potential of the gauge field at spatial infinity,
and on the value of the cosmological constant $\Lambda$.
 The numerical calculations are performed with help of the
package FIDISOL, based on the Newton-Raphson iterative
procedure \cite{FIDI}.
  
Our numerical results indicate that all known axially symmetric 
  YMH solutions admit generalizations in an AdS background.
 Moreover, one can also put a small SAdS black hole inside these configurations.
 Qualitatively, the Higgs field and Yang-Mills field behaviour is very
similar to that corresponding to Minkowski spacetime monopoles.
In particular, we notice a similar shape for the functions
$K_i$, $B_i$ and $\Phi_i$ and also for the energy density. 
 
The most important new feature of the AdS solutions 
is that the magnitude $V_0$ of the electric potential $B_1$
at infinity is no longer restricted.
In an asymptotically Minkowski spacetime, this constant is restricted to 
$V_0\le 1$, $i.e.$ $|A_t|\le |\Phi|$. 
For $V_0>1$ some gauge field functions become oscillating instead of
asymptotically decaying which leads to an infinite mass of the solutions. 
However, in an AdS spacetime, finite energy solutions with arbitrary $V_0$ are allowed,
$i.e.$ there are no limits on the value of the electrostatic potential.

For any value of $\Lambda \leq 0$,
the electric charge $Q_e$ can be read from the asymptotics
of the electric potential $B_1$:
\begin{equation}
\label{Qe}
B_1=V_0-\frac{Q_e}{r}+\dots~~. 
\end{equation}
 After replacing  the asymptotic expressions of the solutions in the general expression (\ref{totalJ}), one finds that
the contribution to the total angular momentum from the boundary integral at infinity can be written as 
\begin{equation}
\label{J2k}
{\cal J}_{\infty}= \left[1+(-1)^m \right ] \frac{n Q_e}{2}.
\end{equation}
 Therefore the total angular momentum   (\ref{totalJ}) of  $m=2k+1$ solutions $i.e.$ with a 
net magnetic charge is given entirely by the contribution of the inner boundary term, $J=-{\cal J}_{(e.h.)}$.
However, that term vanishes as $r_h\to 0$.
As a result, the solutions in an AdS background with $Q_m\neq 0$
have a zero total angular momentum\footnote{Note that these solutions  possess a 
nonvanishing angular momentum density, $T_\varphi^t\neq 0$ (this holds for configurations in 
the Prasad-Sommerfield limit as well).}.
The situation is different for configurations with $m=2k$, $i.e.$ for a vanishing magnetic charge, $Q_m=0$.
The total angular momentum of such solutions in an  AdS spacetime
is proportional to the electric charge, $J=n Q_e$.
However, due to the event horizon contribution to the general relation (\ref{totalJ}),
this simple relation does not hold for their generalization in
a SAdS background. 

\begin{figure}[hbt]
\lbfig{fig:1}
\begin{center}
 \hspace*{-1.24cm}
\includegraphics[height=.2\textheight, angle =0]{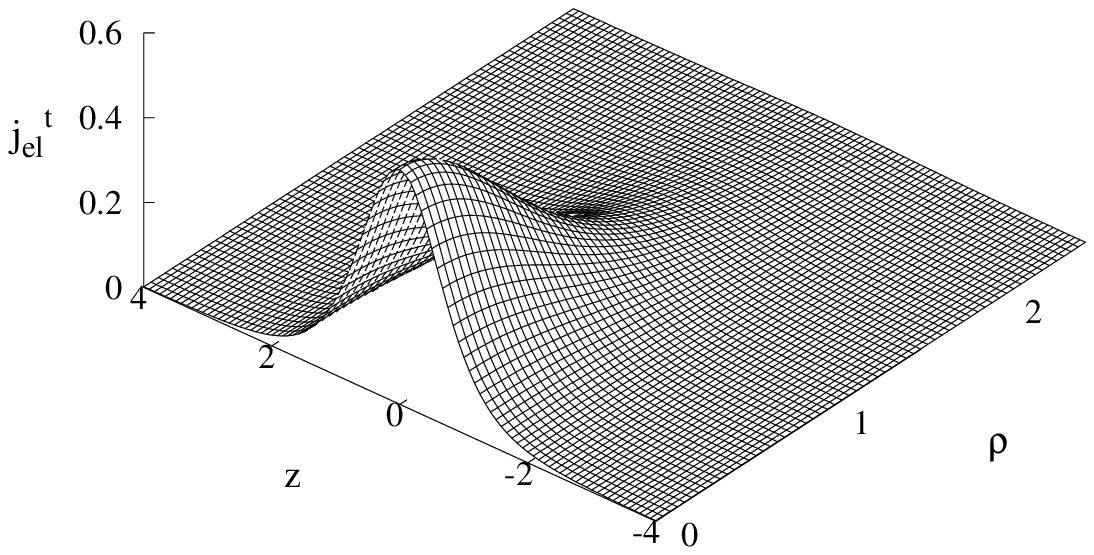}
\includegraphics[height=.2\textheight, angle =0]{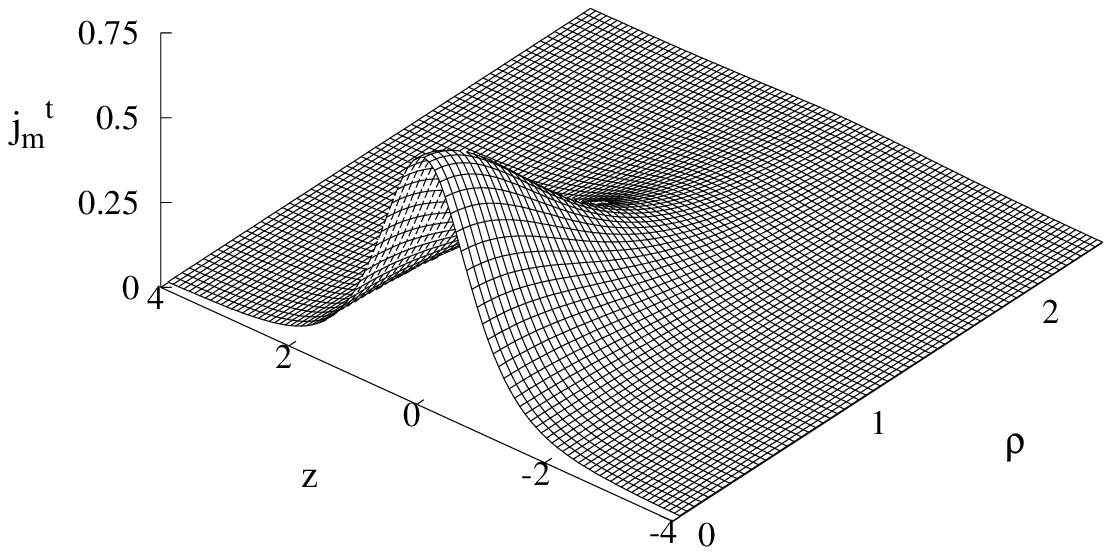}
\includegraphics[height=.12\textheight, angle =-0]{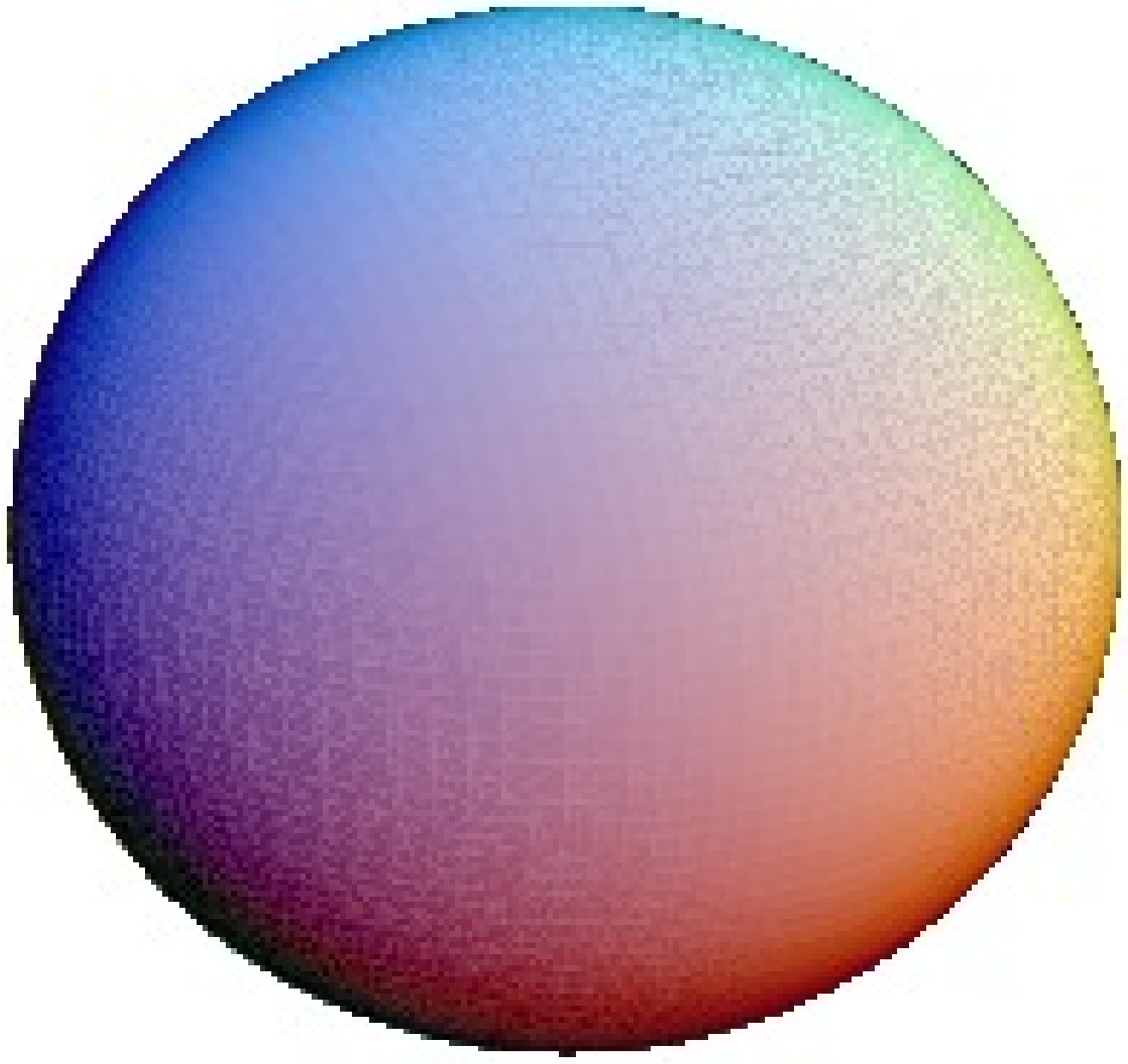}
 \hspace*{-0.2cm}
\includegraphics[height=.2\textheight, angle =0]{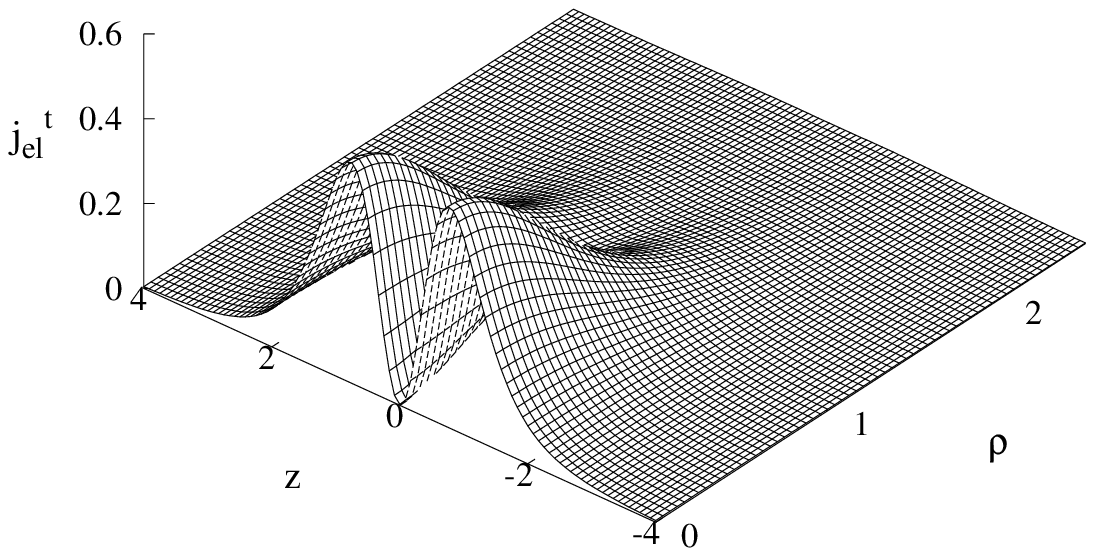}
\includegraphics[height=.2\textheight, angle =0]{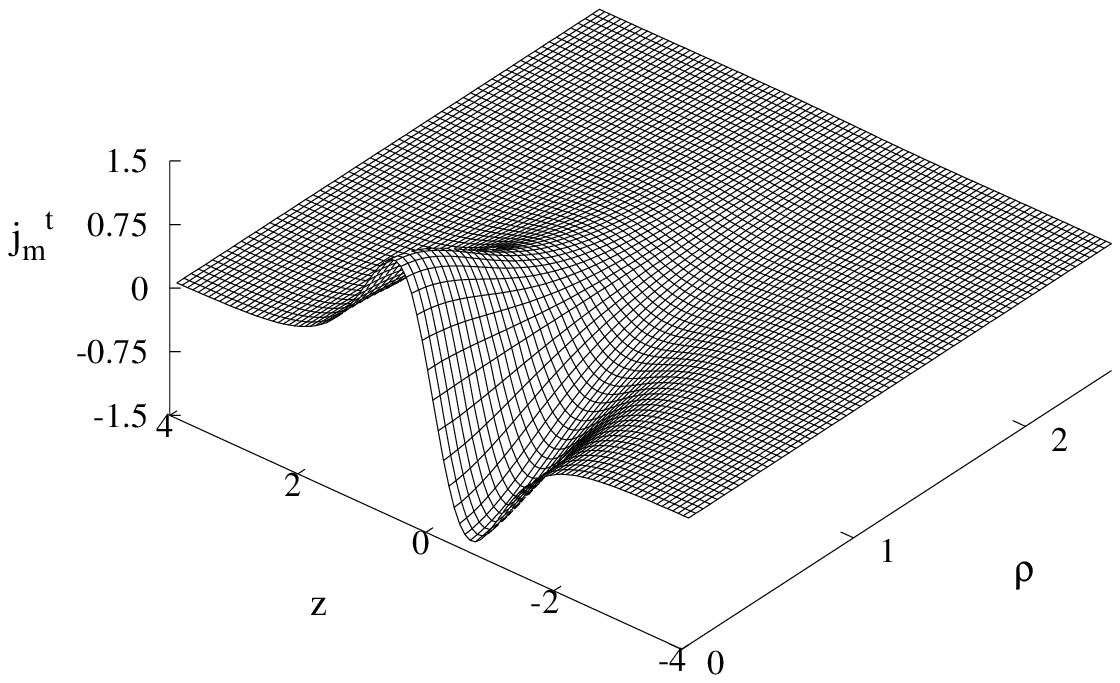}
\includegraphics[height=.15\textheight, angle = 65]{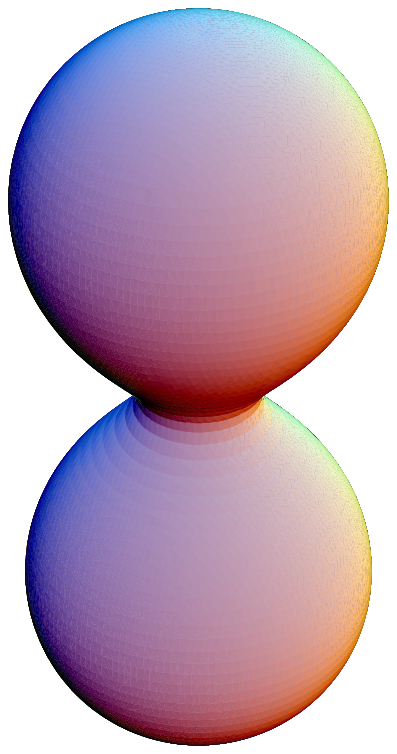}
 \hspace*{-0.2cm}
\includegraphics[height=.2\textheight, angle =0]{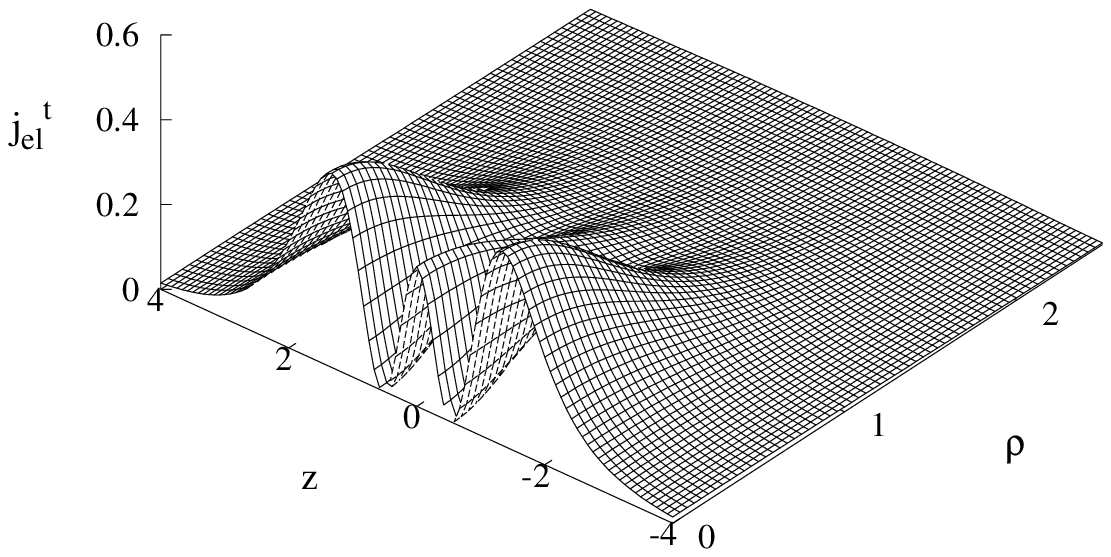}
\includegraphics[height=.2\textheight, angle =0]{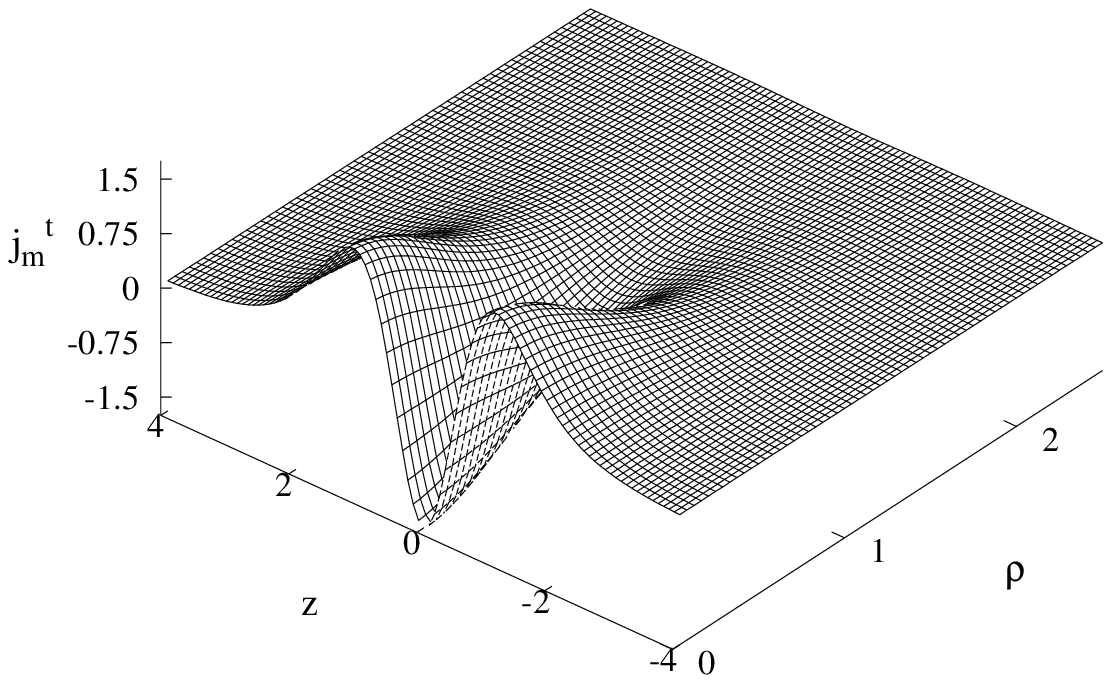}
\includegraphics[height=.16\textheight, angle =65]{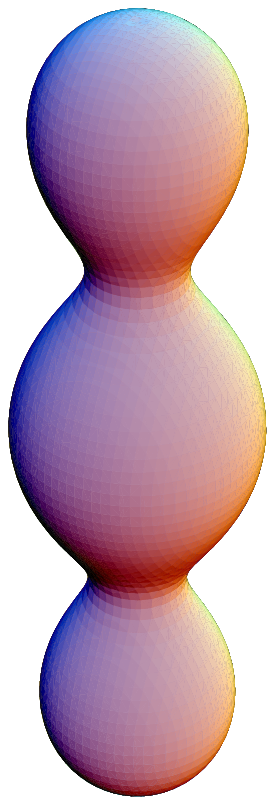}

\end{center}
\vspace{-0.5cm}
\caption{\small The electric charge density (left panels) and the magnetic
charge densities (middle panels) are shown  at $V_0=1, \Lambda=-1/3$ as functions
of the coordinates $\rho = r \sin \theta, z=r\cos \theta$ for:
(i) $n=1, m=1$ spherically symmetric AdS dyon (first set);
(ii) $n=1, m=2$ AdS dyon pair  regular solution (middle set);
(iii) $n=1, m=3$  three dyon chain.
Energy isosurfaces of these configurations are exhibited on the right panels.
}
\end{figure}

\subsection{Axially symmetric solutions in 
globally AdS spacetime}
 
Let us start by recalling that the configurations we are discussing here  are
characterized by two integers, $ m$ and  $n$, where $m$
is related to the polar angle and $n$ to the azimuthal angle.
For both Minkowski and AdS backgrounds, the
regular solutions  of the system (\ref{feqA}) 
correspond to electrically charged (multi)monopoles ($m=1, n\geq 1$) with
magnetic charge $Q_m=n$, while the configurations with $m=2, n=1,2$ correspond to electrically charged monopole-antimonopole
pairs with zero net magnetic charge. 
In general, configurations with
$m \ge 2, n=1,2$ correspond to chains of $m$ monopoles and antimonopoles with a net electric charge.
 Here, the Higgs field vanishes at $m$ isolated points along the symmetry axis $z=r\cos \theta$.

The topological reason for the appearance of these saddle point solutions of the YMH equations
is related to the existence of noncontractible loops in the configuration space of the model,
minimization of the energy functional along such a loop yields an equilibrium state in the middle
of the loop, where the constituents are in stationary equilibrium \cite{Taubes}. 
Since in this case 
we have a complicated pattern of short-range interactions between the constituent monopoles,
it is instructive to use a simplified picture of the effective electromagnetic interaction
between the poles and electric current rings which, according to the equations of the 't Hooft field
tensor \re{Hooft_tensor} both generate the abelian magnetic field which is supporting
equilibrium (see \cite{S06} for a detailed discussion of these aspects).

The results for $\Lambda=0$ show that,
as the winding number $n$ increases further, $i.e.$, $n \ge 3$,
instead of isolated nodes on the symmetry axis, in the flat space limit 
vortex ring solutions arise, where the Higgs field vanishes on one or more rings,
centered around the symmetry axis 
\cite{Kleihaus:2003nj}, 
\cite{Kleihaus:2003xz},
\cite{Kleihaus:2004is}.
However, the coupling of the matter fields to gravity
and/or an increase of the electric charge of the configurations 
\cite{Kleihaus:2004fh},
\cite{Kleihaus:2007vf},
\cite{Kleihaus:2005fs}
can change the
situation drastically, since the structure of the nodes of the Higgs fields
depends strongly on the scalar coupling, the magnitude of the electric charge of the system and on the gravitational constant.
Therefore, the information about the structure of the nodes becomes less important. 
Furthermore, new branches of solutions may appear at 
critical values of the parameters  
\cite{Kleihaus:2007vf},
\cite{Kleihaus:2005fs},
\cite{Kunz:2007jw},
\cite{Kunz:2006ex}.

We have found that all basic features of these nongravitating solutions with $\Lambda=0$ 
repeat in the case of an AdS background. 
As the winding number $n$ increases, the attraction between the constituents increases, 
so for $n\ge 2$ the
energy density distribution is deformed to the system of $m$ tori centered around the symmetry axis.
A negative cosmological constant introduces an attractive 
force, which reduces the typical size of the configurations.
Then, for large enough 
$|\Lambda|$ the additional AdS gravitational attraction makes difficult to 
 classify the individual constituents
according to the position of the nodes of the scalar field as it is possible in the flat space. 

 We illustrate the AdS solutions with a few examples in Figure 1.
There we exhibit the distribution of the electric and magnetic charge densities $j^t_{el},j^t_m$, and the energy isosurfaces for the 
spherically symmetric dyon ($m=n=1$), the two dyon pair ($m=2,n=1$) and chain of 3 dyons ($m=3,n=1$). 
In these solutions the $m$ individual constituents are 
located on the symmetry axis, with roughly equal distance
between them. 
(There we fix the values $\Lambda =-1/3$ and $V_0 =1$ for the cosmological constant and the 
electric potential,
respectively; a similar picture has been found, however, for 
other values of these parameters.)

In Figure 2  we show the mass-energy of $(m=3,4$;~$n=1)$ YMH chain solutions with fixed electric charge (left) and fixed electric
potential (right) as a function of the cosmological constant.
One can notice the existence of AdS solutions with $V_0>1$;
however, as expected, the mass-energy of such configurations
 diverges as $\Lambda \to 0$.
A similar picture has been found for solutions with $m=1$ and $m=2$ 
(see also \cite{vanderBij:2002sq}, \cite{Radu:2004ys}).

\subsection{Axially symmetric solutions in 
the Schwarzschild-AdS background}
 
 According to the standard arguments, one expects these solutions
 to possess generalizations in a SAdS background.
 This is indeed confirmed by the numerical analysis.
However, for all solutions reported in this work there is an upper bound on 
the event horizon radius, typically with $r_h^{(max)}<\ell$.
The possible existence of large SAdS black holes
with YMH hair is an open question and may require
considering a region of the parameter space not covered
by our numerical study ($e.g.$ very large values of the electric charge).

\begin{figure}
\begin{center}
\includegraphics[height=.25\textheight, angle =0]{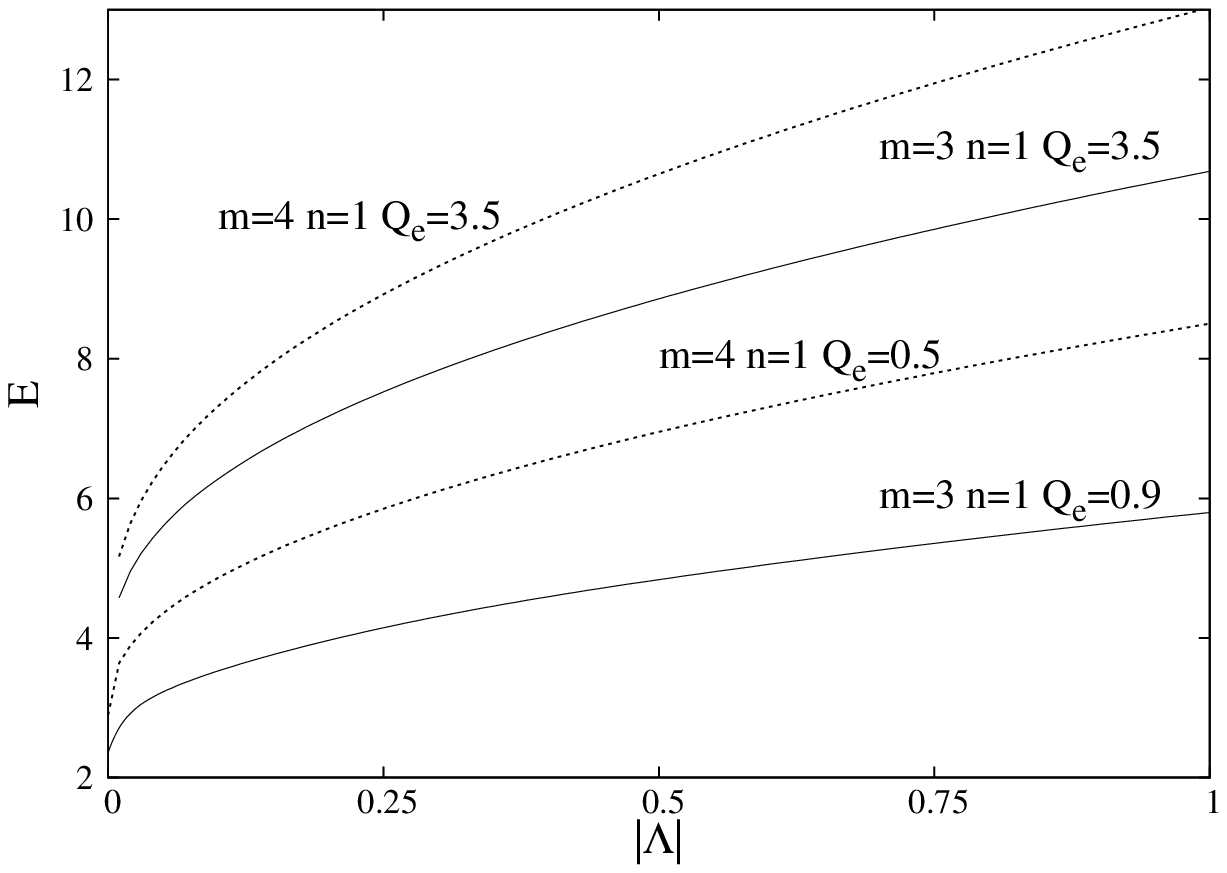}
\includegraphics[height=.25\textheight, angle =0]{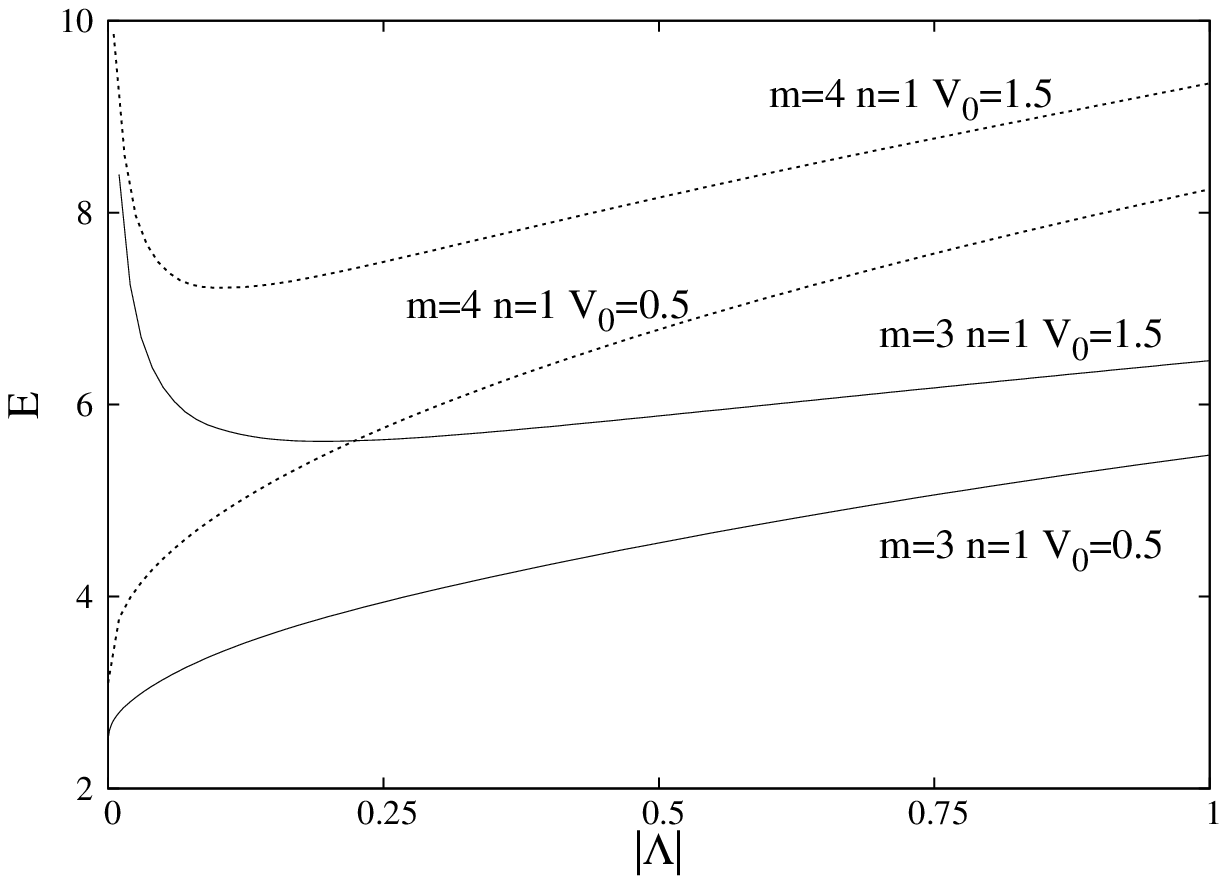} 
\end{center}

\caption{\label{fig:2}\small  The mass-energy $E$ is shown as a function of the cosmological constant
for $m=3,4$ AdS YMH chains with a fixed electric charge (left) and a fixed electric potential (right).}
\end{figure}

\begin{figure}
\begin{center}
\includegraphics[height=.25\textheight, angle =0]{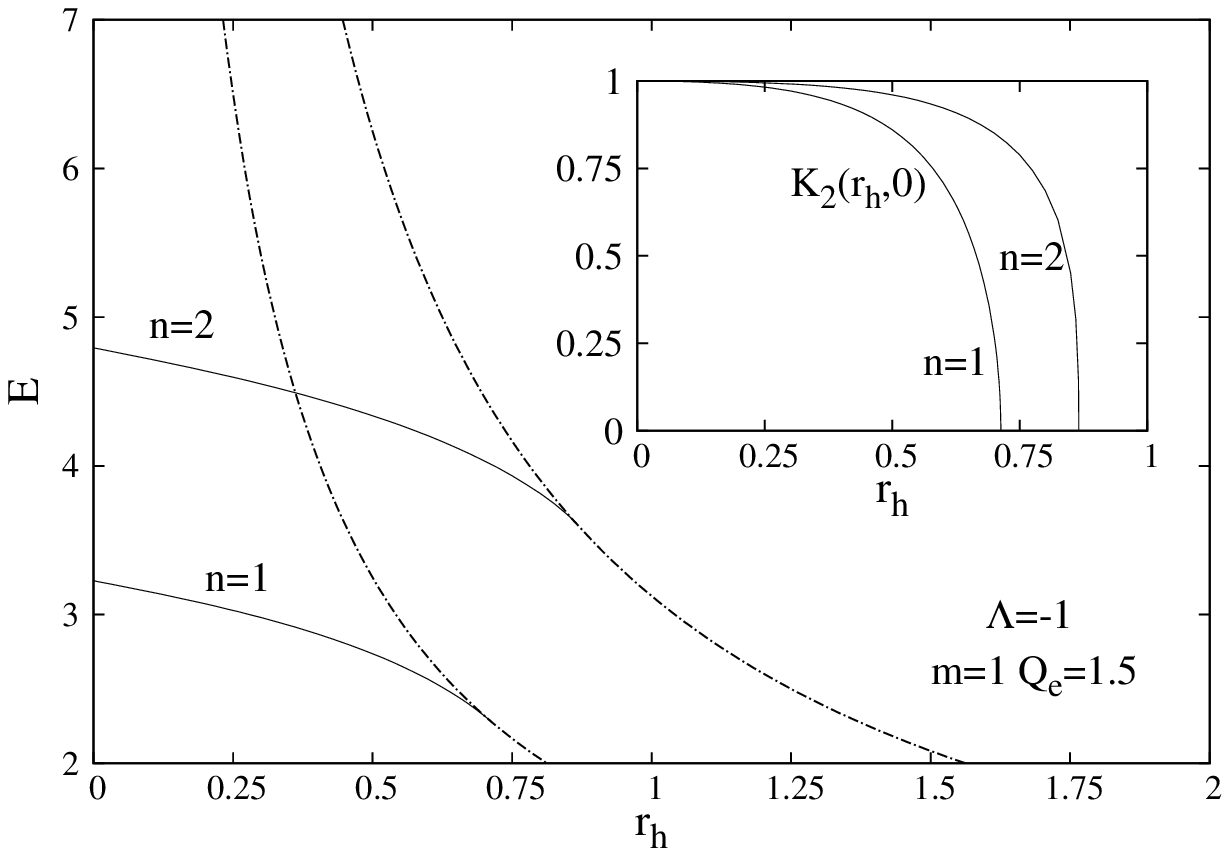}
\includegraphics[height=.25\textheight, angle =0]{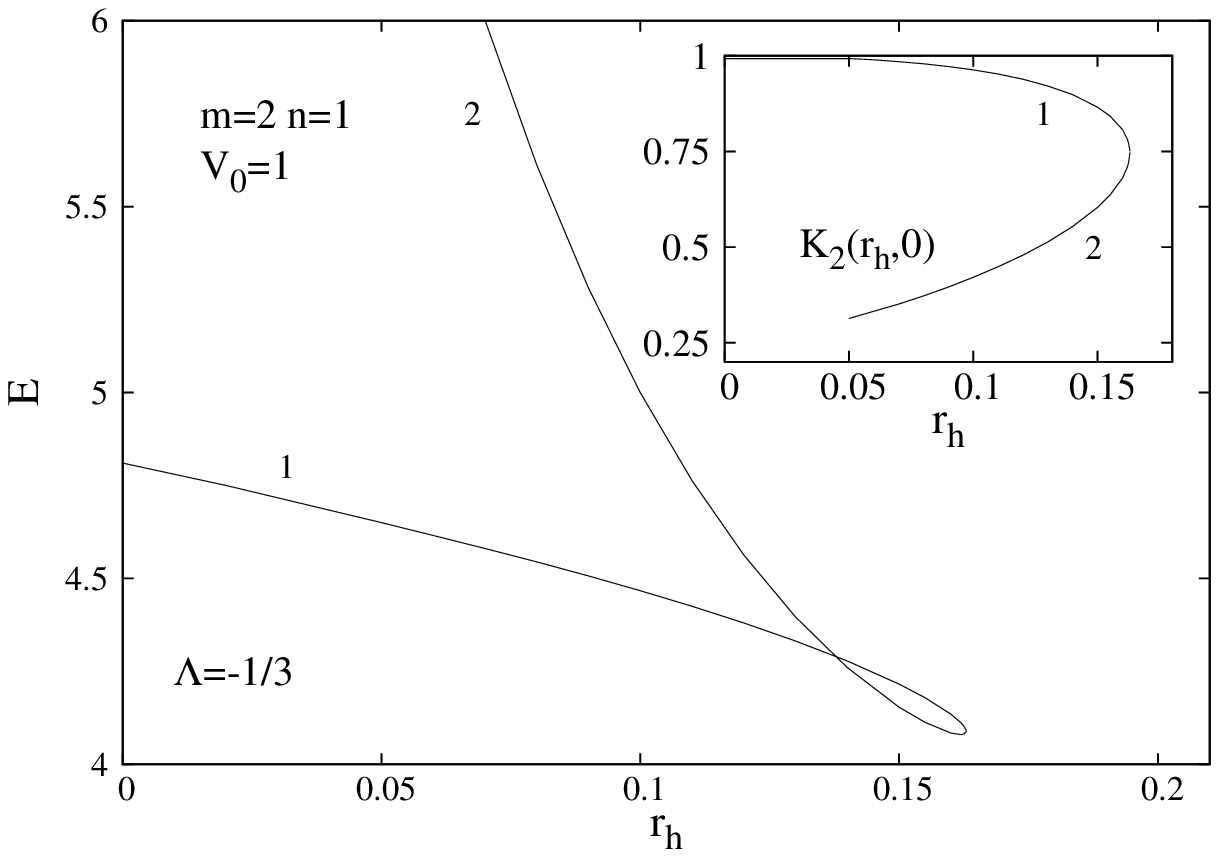} 
\end{center}

\caption{\label{fig:3}\small  The mass-energy $E$ is shown 
as a function of the
event horizon radius $r_h$ for electrically charged YMH solutions in a fixed Schwarzschild-AdS background
with a net magnetic charge, $m=1$ (left) and with a vanishing total magnetic charge, $m=2$ (right).
The insets show the value of the magnetic gauge potential $K_2$ at $\theta=0$
on the horizon.
For $m=1$, the branch of Abelian solutions as given by (\ref{sol1}) is also shown (the dotted curves). 
}
\end{figure}

Let us start with a discussion of the fundamental configurations with $m=1$.
In this case, an  important role is played by the 
 Abelian dyon solution (with $V_0=Q_e/r_h$)
 \be
\label{sol1}
K_1=K_2=K_3=K_4= 0,~
B_1 =V_0-\frac{Q_e}{r},~~B_2=0,~~
\Phi_1= 1,~~\Phi_2 = 0 .
\ee
This solution exists for all values of $r_h$
and has a finite mass-energy $E= (n^2+Q_e^2)/2r_h$ and a vanishing angular momentum $J=0$.
 
We have found numerical evidence that, for a given value of the electric charge,
a branch of $m=1,n\geq 1$ non-Abelian solutions bifurcates from the Abelian configuration (\ref{sol1}),
for a critical value of the event horizon radius $r_h$.
Close to the bifurcating point, the YM potentials and the Higgs fields are written as a sum
of the  solution (\ref{sol1}) and a small perturbation $\delta F=(\delta K_i$, $\delta \Phi_i,\delta B_i)$.
Linearizing the YMH equations with respect to these perturbations leads to an eigenvalue
matrix equation for $r_h$  of the form
\be
\label{eq1}
\left ( (\partial_{rr}+\frac{1}{r^2N(r)}\partial_{\theta \theta})I_{8\times 8}
+M_r \partial_{r}+M_\theta \partial_{\theta} +M_0 \right )\delta F=0,
 \ee
with $M_r$, $M_\theta$ and $M_0$ are $8\times 8$ matrices, depending
only on the Abelian solution (\ref{sol1}) and the metric function $N(r)$, with a complicated expression.
The above equation simplifies only for the spherically symmetric solutions,
in which case $\delta F=0$ (the higher order terms, however, are not zero) 
except for $\delta K_2=\delta K_4=w_1(r)$, which solve the equation
\be
\label{eq2}
w_1''
+\frac{N'}{N}w_1'
+\frac{1}{r^2 N}\left(1-r^2+\frac{r^2 (\frac{Q_e}{r_h}-\frac{Q_e}{r})^2}{N} \right)w_1
=0,
 \ee
 with the boundary conditions $w_1(r_h)=b>0$, $w_1(\infty)=0$.
 The critical values  of $r_h$ found in this way for given electric charges (or, equivalently, for 
  given electrostatic potentials $V_0$) are in very good agreement with those found 
by directly solving the set of YMH equations.

The respective branch of non-Abelian solutions continues inwards in $r_h$, joining smoothly for $r_h=0$
the corresponding $(m=1,n)$ dyonic solution in a fixed AdS background
discussed in the previous sub-Section. 
As expected, along this branch, the mass of YMH solutions is smaller than the mass of the corresponding
Abelian configurations, $i.e.$
they are thermodynamically favoured\footnote{Here we are comparing solutions in a canonical ensemble, $i.e$ with the same 
electric charge.}. This feature appears to be universal, being recovered for all considered values of 
$\Lambda$. 

To illustrate this behaviour,  we exhibit in Figure 3 (left) the mass-energy
 and the value of magnetic gauge potential $K_2$  at $r=r_h$, $\theta=0$ 
for $m=1$ dyons with  $n=1,2$  
versus the horizon radius $r_h$ 
(these results are found for fixed values of the electric charge
and cosmological constant, $Q_e=1.5$ and $\Lambda=-1$, respectively). 
The dotted curve  there corresponds to the branch of Abelian solutions with the same values of $Q_e,n$ and $r_h$.
One can see that a fundamental 
branch of dyons emerges from the corresponding AdS
solution with $r_h=0$ and extends up to a maximal value
of the horizon radis $r_h^{(max)}$ where it merges with the Abelian branch. 
Similar results are found when studying instead solutions with a fixed electric potential $V_0$
($i.e.$ in a grand canonical ensemble),
the YMH configurations emerging again as perturbations
of the Abelian solution (\ref{sol1}). 

A different picture was found for solutions with $m=2$, $i.e.$
electrically charged monopole-antimonopole pairs.
In this case there are no branches of YMH solutions 
 emerging as perturbations of a (electrically charged and magnetically neutral)
critical Abelian configuration.
As a result,
 one finds a different dependence on the horizon radius, which is
 illustrated in Figure 3 (right). 
%
Again, a lower branch of non-Abelian 
solutions (label `1' in Figure 3 (right)) emerges in the limit of small $r_h$ from the regular AdS chain configuration. 
This branch extends in $r_h$ and bifurcates
with a second  branch at some maximal value of the horizon radius (for example, for the $n=1$ pair
of dyons with $V_0=1,~\Lambda=-1/3$, one finds $r_h^{(max)} \simeq 0.16$). 
This second branch extends backwards in $r_h$.
Since the upper 2-dyon branch is axially symmetric, it is not linked
to the trivial Abelian branch anymore; instead, as the horizon radius decreases to zero, it 
appears to approach a 
solution of the Bartnik-McKinnon type \cite{Volkov:1998cc},
$i.e.$ with a trivial Higgs field
and nonvanishing non-Abelian potentials 
(however, this upper limit is rather difficult 
to study numerically).
 Note also that the mass of these solutions in a fixed SAdS background 
exibits a loop close to the $r_h^{(max)}$,
 when plotted in terms of $r_h$.
Outside the loop, the second branch possesses a higher mass than the
 first branch. 

We have also constructed solutions with  $m=3,4$ and $n=1,2,3$,
although with a lower numerical accuracy.
Not completely unexpected, these solutions follow the pattern
found for the $m=2$ case above.
Working again with a fixed electric potential $V_0$,
there is always a lower branch of configurations smoothly emerging from the 
corresponding $(m,n)$ solutions in a fixed AdS background,
which extends up to a maximal value of the horizon radius $r_h$.
There it joins a secondary branch of solutions, which extends
backwards in $r_h$.

The reason why the $m\geq 2$ solutions show a different pattern can 
be understood heuristically by noticing that they are composite, saddle point
configurations.
Thus, they are unstable and we cannot expect a 
(single component) Abelian solution to decay into them.


\section{Further remarks. Conclusions }
We have given numerical evidence that,
when the global AdS spacetime 
replaces Minkowski spacetime as the background geometry,
all known YMH axially symmetric solutions admit generalizations
with rather similar properties.
However, a more complicated picture emerges when the solutions are studied
in a fixed SAdS black hole background, 
the configurations  with the lowest windining number $m=1$
emerging as perturbation of some critical Abelian dyonic solutions.

One may ask if these solutions play some role in the conjectured AdS/CFT correspondence \cite{Maldacena:1997re}.
Indeed, the ($m=1,n=1$) spherically symmetric dyonic solutions of the YMH model
have found an interesting interpretation in this context 
\cite{Lugo:2010qq}, 
\cite{Lugo:2010ch}, 
\cite{Allahbakhshi:2010ii},
\cite{Allahbakhshi:2011nh}.

Since we have not started with a consistent truncation
of string theory,
we do not have a detailed microscopic description of the  
dual theory for the YMH action (\ref{model}).
Nevertheless, some basic elements elements of the gravity/gauge duality dictionary \cite{Maldacena:1997re}
still allow us to say the following.
First, the dual theory is defined in an Einstein universe in $d=3$ dimensions, with a line element
$ds^2=\ell^2(d \theta^2+\sin^2 \theta d \varphi^2)-dt ^2$.
 In this approach, the Hawking temperature
of the SAdS black hole corresponds to the temperature of the $d = 3$ system.
Also,
the $SU(2)$ gauge
symmetry of the bulk action corresponds 
to a global $SU(2)$ symmetry in the dual field theory.
As usual in models with an electric field,
the chemical potential and the electric charge density of the 3-dimensional system
are defined from the asymptotics (\ref{Qe}) of the bulk electric gauge potential, 
the charge density operator being proportional to $Q_e$.
Moreover, we have seen that, for odd values of $m$ and any winding number $n$,
the magnetic gauge potential does not trivialize as $r\to \infty$.
In an AdS/CFT context, this boundary value 
 plays the role of a magnetic source for the dual field theory. 
 Concerning the Higgs field, we notice that 
for the solutions in this work without a scalar potential $\lambda=0$, 
its generic asymptotic behaviour is
\begin{equation}
\Phi^{(a)}=v^{(a)}+\frac{\beta^{(a)}}{r^3}+\dots,
\end{equation}
with $v^{(a)}$ a source for a triplet of operators $\Psi^{(a)}$ in the dual theory.
However, we can always choose a gauge such that $v^{(a)}$ is constant on the boundary.
In such a gauge the magnetic field on the boundary corresponds to the field of vortices. 
For example the axially symmetric multimonopole corresponds to a vortex of magnetic flux $nQ_m$
whereas the monopole-antimonopole pair in the bulk 
corresponds to system of vortices with opposite fluxes. 
Therefore the transition between Abelian and non-Abelian
branches in the bulk at some critical value of the horizon radius, $i.e.$ Hawking temperature, is considered to be 
dual to the phase transition  on the boundary, related with the condensation of vortices.   

As avenues for further research, it would be interesting to extend the solutions in this work
by including the effects of the back reaction on the spacetime geometry, 
especially in the black hole case.
Here we expect that some features revealed in Section 3, in particular
the existence of an instability of the Abelian dyon configurations, will remain valid,
translating into instabilities (and corresponding new non-Abelian branches) of the 
Kerr-Newman--AdS black hole.
Another interesting possible extension of our solutions would be to construct
their counterparts for a Poincar\'e patch of AdS, in which case the dual theory will be defined in a 
boundary metric 
corresponding to $2+1$ dimensional Minskowski spacetime.

\vspace{0.65cm}
\noindent{\textbf{~~~Acknowledgements.--~}} 
We would like to thank Burkhard Kleihaus for useful discussions.
O. Kichakova, J. Kunz and E. Radu gratefully acknowledge support by the DFG,
in particular, also within the DFG Research
Training Group 1620 ''Models of Gravity''.
The work of Ya. Shnir  is supported by the Alexander von Humboldt Foundation.   


\end{document}